\documentclass[final,5p,times,twocolumn]{elsarticle}

\usepackage{amssymb}
\usepackage[pdftex]{color}

\begin{document}

\begin{frontmatter}



\title{History-dependent deformation of a rotated granular pile governed by granular friction}


\author[N1]{Terunori Irie}
\author[N2]{Ryusei Yamaguchi}
\author[N1]{Sei-ichiro Watanabe}
\author[O1]{Hiroaki Katsuragi}
  
  \address[N1]{Department of Earth and Environmental Sciences, Nagoya University, Furocho, Chikusa, Nagoya 464-8601, Japan}
  \address[N2]{Technical Center, Nagoya University, Furocho, Chikusa, Nagoya 464-8601, Japan}
  \address[O1]{Department of Earth and Space Science, Osaka University, 1-1 Machikaneyama, Toyonaka 560-0043, Japan}

  
\begin{abstract}
We experimentally examined the history dependence of the rotation-induced granular deformation. As an initial state, we prepared a quasi-two-dimensional granular pile whose apex is at the rotational axis and its initial inclination is at the angle of repose. The rotation rate was increased from $0$ to $620$~(rpm) and then decreased back to $0$. During the rotation, deformation of the rotated granular pile was captured by a camera. From the acquired image data, granular friction coefficient $\mu$ was measured as a function of the ratio between centrifugal force and gravity, $\Gamma$. To systematically evaluate the variation of $\mu$ both in the increasing (spinning up) and decreasing (spinning down) rotation-rate regimes, surface profiles of the deformed granular piles were fitted to a model considering the force balance among gravity, friction, and centrifugal force at the surface. We found that $\mu$ value grows in the increasing $\Gamma$ regime. However, when $\Gamma$ was reduced, $\mu$ cannot recover its initial value. 
 A part of the history-dependent behaviors of the rotated granular pile can be understood by the force balance model. 
\end{abstract}



\begin{keyword}


  Centrifugal force \sep granular friction \sep hysteresis \sep force-balance model

\end{keyword}
\end{frontmatter}


\section{Introduction}
\label{sec:intro}
Due to the complex nature of nonlinear interactions among grains, behaviors of granular matter frequently become nontrivial~\cite{Duran:2000,Aranson:2009,Andreotti:2013}. Such nontrivial granular behaviors have attracted various scientists and engineers. As a consequence, fundamental understanding of granular matter has been developed in recent decades. The proper understanding of granular behavior is certainly important for various industrial applications as well. To effectively handle granular materials, we have to control the rheological properties of granular matter~\cite{Nedderman:1992,Rao:2008}. Therefore, granular behaviors under the various conditions have been studied so far. For example, mixture of water and granular matter is crucial in various applications~\cite{Herminghaus:2013}. For dilute granular gas, kinetic theory has been used to understand its behaviors~\cite{Brilliantov:2004}. The vibrated dense granular matter exhibiting various phenomena has also been studied extensively~\cite{Rosato:2020}. These recent progresses of granular characterization enable us to efficiently handle granular materials in some situations. However, our understanding of various granular behaviors is still very limited. There are many unsolved counterintuitive phenomena produced by granular matter. An example of such complex granular behaviors is its history dependence due to the nonlinearity of granular mechanics. The history dependence of granular behavior is the main focus of this study.

Physical understanding of granular behaviors is necessary also for solving planetary-related problems. Because most of the terrestrial bodies in the solar system are covered with granular matter (regolith/boulders), planetary surface processes are governed by physical behaviors of granular matter~\cite{Melosh:2011}. Thus, physical properties of granular matter are extensively studied recently in the field of planetary science. For instance, low-speed impact on a granular target layer was studied to collect fundamental data for planning touchdown missions on asteroidal surfaces~\cite{Ballouz:2021,Thuillet:2021,Sunday:2021}. Low-speed granular impact has been investigated also by granular physicists~\cite{Ruiz:2013,Katsuragi:2016,vanderMeer:2017}. 
To understand the penetration dynamics into a loose granular layer, frictional drag term has been a key factor~\cite{Katsuragi:2007}. Recently, the depth-dependent pseudo frictional force was modeled by granular Archimedes' principle~\cite{Kang:2018}. Surface processes occurring on the surface of small bodies can also be governed by granular friction. Both by observation and  numerical simulation, crucial roles of frictional behavior governing global shape and surface terrain of asteroids have been investigated~\cite{Watanabe:2019,Sugiura:2021}. 

While granular friction is an important key to understand various granular-related behaviors, its properties are quite complex. In dense flow regime, phenomenological granular friction laws have been established~\cite{GDRMiDi:2004,daCruz:2005,Jop:2006,Pouliquen:2006}. However, applicabile range of these friction laws is limited to the flowing regime.

In static situations, granular friction can be characterized by the angle of a stable granular heap~\cite{Nagel:1992}. When we continuously drop a lot of grains on a flat floor, a conical granular heap is developed. The angle of this conical shape is called angle of repose, $\theta_r$. The angle of repose can be related to the phenomenological friction coefficient $\mu$ as, $\tan\theta_r = \mu$. Because the angle of repose is necessary information to evaluate the landscapes on various planetary bodies, its gravity dependence has been intensively studied in geophysical and planetary sciences. For instance, rotating drums have been used to measure the angle of repose under various gravitational conditions~\cite{Klein:1990,Arndt:2006,Cosby:2009,Kleinhans:2011}. In addition, some other methodologies have also been used to characterize the gravity dependence of the angle of repose~\cite{Hofmeister:2009,Blum:2010, Marshall:2018,Chen:2019}. Unfortunately, these previous works are not completely consistent with each other. Some of them reported the gravity-dependent angle of repose~\cite{Klein:1990,Cosby:2009} and the others reported that angle of repose is almost independent of gravity~\cite{Marshall:2018,Chen:2019}. Even the complex gravity dependence of angle of repose was also reported~\cite{Hofmeister:2009,Kleinhans:2011}. Furthermore, in spite of these efforts, precise measurement of the gravity dependence of the angle of repose has been difficult. 

To effectively control the gravity (body force) effect, we can use centrifugal force. In a rapidly rotated situation, centrifugal force can overcome the effect of gravity. Based on this idea, centrifugal force has been used in various granular-related technologies, e.g., pumping~\cite{Hong:2016}, dehydration~\cite{Bizard:2013}, silo flow~\cite{Dorbolo:2013}, and soil handling~\cite{Cabrera:2017}. For the scientific purpose, granular cohesion measurement has also been performed using a rotating system~\cite{Nagaashi:2018,Nagaashi:2021}. Recently, we developed a new apparatus by which the deformation of a quasi-two-dimensional (2D) granular pile can be precisely measured~\cite{Irie:2021a,Irie:2021b}. This type of rotating 2D setup is useful to characterize mechanical properties of granular matter~\cite{Castellanos:2007,SoriaHoyo:2008,Huang:2021}. 

In Refs.~\cite{Irie:2021a,Irie:2021b}, we studied granular friction and cohesion using the newly developed system. By considering the force balance among gravity, friction, and centrifugal force, we developed a simple analytic model for the surface deformation of the rotated granular pile. The model can explain the experimental data very well. Using the model, we revealed the body-force dependence of the granular friction~\cite{Irie:2021b}. We also measured the effective cohesion strength of the rotated granular pile~\cite{Irie:2021a}. Local angle distribution was also measured by Huang et al. using a similar experimental setup~\cite{Huang:2021}. 

In these studies, however, history dependence of the deformation of granular piles has not been analyzed. In general, granular behaviors can be history dependent. The deformation and friction of granular matter might depend on the history of loading. Granular friction and its protocol dependence are very important for various chemical engineering processes~\cite{Rao:2008,Fayed:1997}. For efficiently handling granular materials in various industrial applications, history-dependent nature must be understood at least empirically. In addition, history-dependent granular friction and its gravity dependence are crucial also for some scientific studies. Particularly, gravity dependence of angle of repose is crucial to consider the terrain dynamics occurring on the surfaces of planetary bodies~\cite{Klein:1990,Cosby:2009,Kleinhans:2011,Hofmeister:2009,Kawamura:2012}. Therefore, we focus on the history dependence of granular deformation and friction in the rotated granular pile, in this study. This fundamental study relates to a wide range of industrial and scientific investigations.

\section{Experiment}
\begin{figure}
\centering
\includegraphics[clip,width=0.6\linewidth]{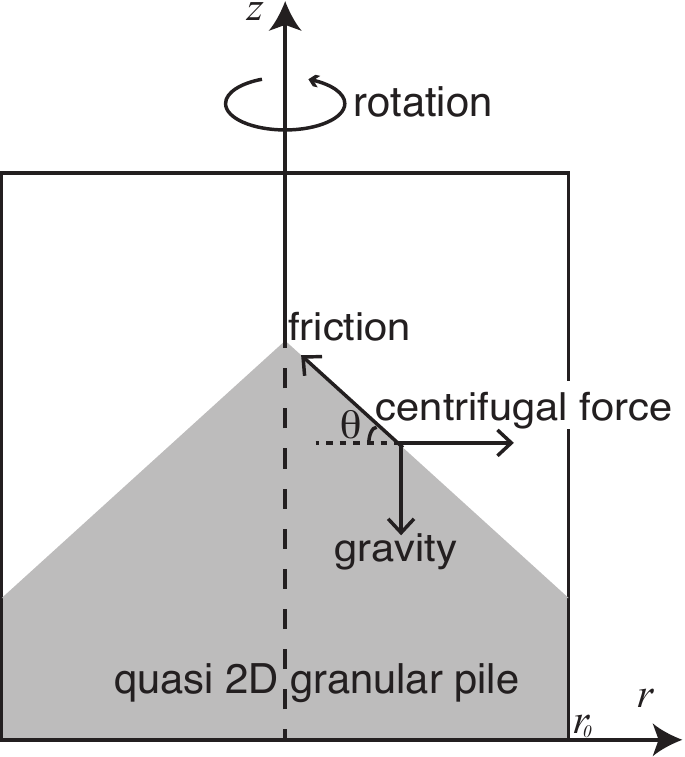}
\caption{
Experimental setup, coordinate system, and relevant forces are schematically shown. A quasi-two-dimensional granular pile was rotated around the central (vertical) axis. The local shape of the granular pile should be determined by the balance between gravity, friction, and centrifugal force. The local slope of the granular pile $\theta$ is defined as shown in the figure. Clockwise direction is defined as positive. The initial value of $\theta$ is certainly the angle of repose $\theta_r$ which corresponds to the critically stable shape in terms of the force balance.
}
\label{fig:setup}
\end{figure}

The experimental apparatus we used in this study is identical to that used in Refs.~\cite{Irie:2021a,Irie:2021b}. Here, we briefly summarize the experimental setup and conditions. We prepared a granular pile with the angle of repose $\theta_r$ in a quasi-2D cell (inner dimensions: 100~mm $\times$ 100~mm $\times$ 10~mm). The front wall is made of a transparent acrylic wall to directly observe the deformation. The apex of the pile is located at the center of the cell which corresponds to the rotation axis. The vertically upward direction and horizontal direction in the 2D cell correspond to $z$ and $r$ axes, respectively~(Fig.~\ref{fig:setup}). To quantify the rotation state, we introduce a dimensionless parameter, 
\begin{equation}
  \Gamma = \frac{r_0 \omega^2}{g},
  \label{eq:Gamma_def}
\end{equation}
where $r_0=50$~mm, $\omega$, and $g=9.8$~m~s$^{-2}$ are the half of cell width (radius), rotational angular speed, and magnitude of Earth's gravitational acceleration, respectively. This dimensionless parameter indicates the relative strength of the centrifugal force to the gravitational force at $r=r_0$. In this experiment, we stepwisely increased $\Gamma$ from $0$ to $22$ (corresponding to $620$~rpm rotation rate). After that, $\Gamma$ was stepwisely decreased from $22$ to $0$. Pictures of the deformed granular piles were taken in both increasing and decreasing $\Gamma$ regimes. These pictures correspond to the raw data obtained in this experiment. In our previous works~\cite{Irie:2021a,Irie:2021b}, we have confirmed that gradual deformation induced by this protocol proceeds quasi-statically. Moreover, another similar experiment has also reported quasi-static deformation of the rotated granular pile~\cite{Huang:2021}.

\begin{figure}
\centering
\includegraphics[clip,width=1.0\linewidth]{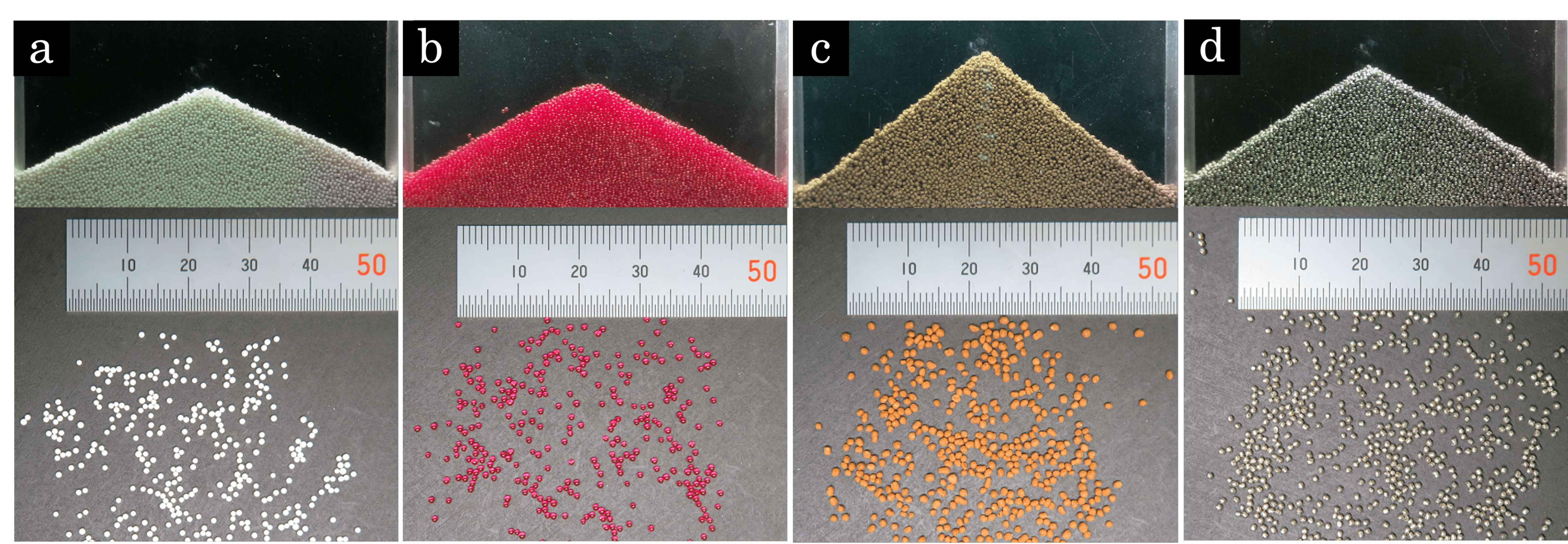}
  \caption{Pictures of grains used in this study, (a)~alumina beads, (b)~colored glass bead, (c)~fish-feed particles, and (d)~cut-wire particles are shown. The upper and lower rows show the piles and grains, respectively.
  }
\label{fig:grains}
\end{figure}

We used alumina beads, colored glass beads (blue, green, red), fish-feed grains, and (stainless-steel) cut-wire grains for constituent particles. Actual grain images are shown in Fig.~\ref{fig:grains}. Grain diameter of all these grains is approximately $d=1$~mm. When the grain size is smaller than $1$~mm, grains are stuck on the front acrylic wall due to the electrostatic force. Thus, we fix $d=1$~mm in this experiment. To prevent electrostatic adhesion, static protection was spayed on the surface of the acrylic plate. Bulk densities of alumina beads, glass beads, fish feed, and cut wire are 2.2, 1.4, 0.61, and 4.6 ($\times 10^3$~kg~m$^{-3}$), and the corresponding angle of repose is $25$, $27$, $35$, and $37$($^{\circ}$), respectively. Basically, these grains possess more or less spherical shape, but the cut wire grains have slightly rough shape. Note that the physical characteristics of differently colored (red, green, or blue) glass beads are same. Each color is used only for identification of the layer of beads in the initial pile~(Fig.~\ref{fig:raw_imgs}(b)). 

Initial granular pile with the angle of repose was prepared by using a funnel except the colored glass beads case. The angle of repose in this setup is defined by the the initial angle of the granular pile formed by using a funnel. We used a mold to form a stripe-patterned granular pile with colored glass beads. Then, these piles were deformed by rotation. To check the reproducibility, we performed three experimental runs with the identical conditions. The data shown in this paper are the average of these experimental runs. Further details about the experimental apparatus and procedures can be found in Refs.~\cite{Irie:2021a,Irie:2021b}.

\section{Results and analyses}
\subsection{Raw images}
\label{sec:raw_images}

\begin{figure*}
\centering
\includegraphics[clip,width=0.8\linewidth]{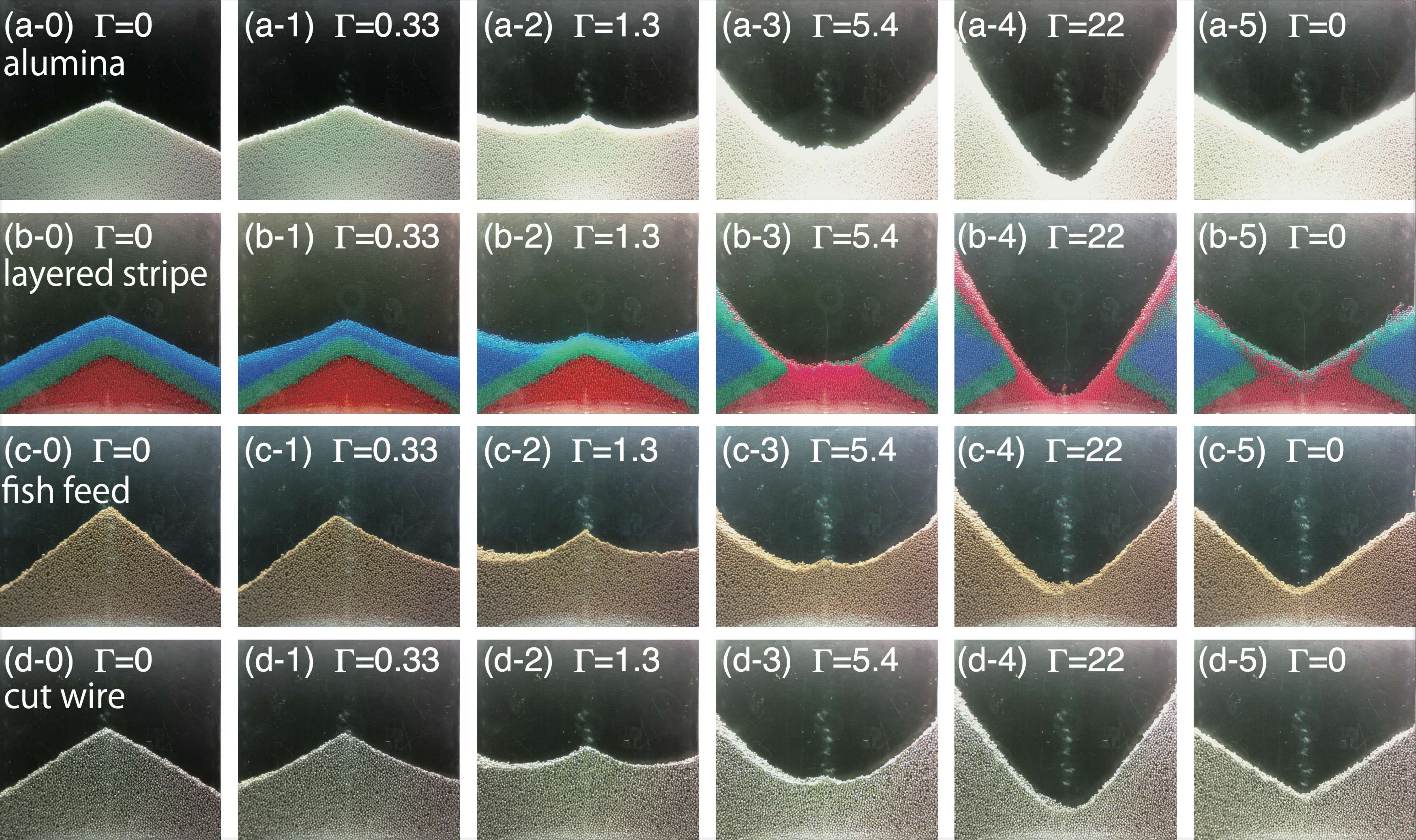}
\caption{
Raw image data of the granular pile deformation. Granular piles consisting of (a) alumina beads, (b) colored glass beads (with layered-stripe structure), (c) fish-feed particles, and (d) cut-wire particles are shown. The left column shows the initial shape, and the right column displays the final shape. Corresponding $\Gamma$ conditions are denoted in each panel. 
}
\label{fig:raw_imgs}
\end{figure*}

The obtained raw data of the granular-pile deformation are shown in Fig.~\ref{fig:raw_imgs}. In Fig.~\ref{fig:raw_imgs}, deformation of granular piles proceeds from left to right. As can be observed, all granular piles show similar deformation behaviors whereas their physical characteristics are different. Moreover, the observed deformation is qualitatively similar to that reported in our previous papers~\cite{Irie:2021a,Irie:2021b}. In these papers, however, deformation of granular piles was investigated only in the increasing $\Gamma$ regime. In Fig.~\ref{fig:raw_imgs}, the final states of the granular piles (after decreasing $\Gamma$) are shown in the right column. As seen in the figure, the final states ($\Gamma=0$) are much closer to the maximum $\Gamma(=22)$ state than the initial $\Gamma=0$ states. This history-dependent deformation is not very surprising. Because the centrifugal force acts to the radially outward direction, the granular piles developed on the side walls are stabler than the initial (centered) granular pile structure. As mentioned later, the surface profiles produced by the maximum $\Gamma$ were kept unchanged when $\Gamma\gtrsim 1$  during the decreasing $\Gamma$ regime. This means that the profile starts collapsing when the centrifugal effect is comparable with the gravity in the decreasing $\Gamma$ regime. Note that, in the increasing $\Gamma$ regime, deformation of granular piles occurs in wider $\Gamma$ regime ($\Gamma \gtrsim 0.1$, Fig.~\ref{fig:raw_imgs})~\cite{Irie:2021a,Irie:2021b}. 

In Fig.~\ref{fig:raw_imgs}(b), deformation of the glass-bead pile with layered stripe is presented. By careful inspection of the images, one can confirm that the deformation is mainly driven by the flow of a thin surface layer. Thickness of this thin layer was measured in \cite{Irie:2021b} and the typical thickness of the fluidized layer is about 5 grain-diameter scale at $\Gamma=22$. Glass beads around the central region are gradually transported to the outward region in the increasing $\Gamma$ regime. And they flow back to the central part when $\Gamma$ is sufficiently decreased. Because different kinds of granular piles show very similar deformation behaviors, we consider the same mechanism governs the deformation of all granular piles. 

While almost all the deformations shown in Fig.~\ref{fig:raw_imgs} look symmetric and qualitatively similar, a certain asymmetry can be confirmed at some instance (e.g. Fig.~\ref{fig:raw_imgs}(c-1)). Because the surface deformation sometimes occurs alternately (not simultaneously), such asymmetry can be observed. To evaluate the average behavior, we will compute the mean profiles by averaging six profiles (left and right of three experimental runs) for each experimental condition. The averaged profiles will be analyzed in the following sections. In Fig.~\ref{fig:raw_imgs}, one can observe that small grains stuck on the central region of the front acrylic plate. However, these stuck grains are much smaller than the typical grain size $d=1$~mm. Namely they are negligible dusty powder.

\subsection{Surface profiles in the increasing $\Gamma$ regime}
\label{sec:profiles_increasing_G}

\begin{figure*}
\centering
\includegraphics[clip,width=0.8\linewidth]{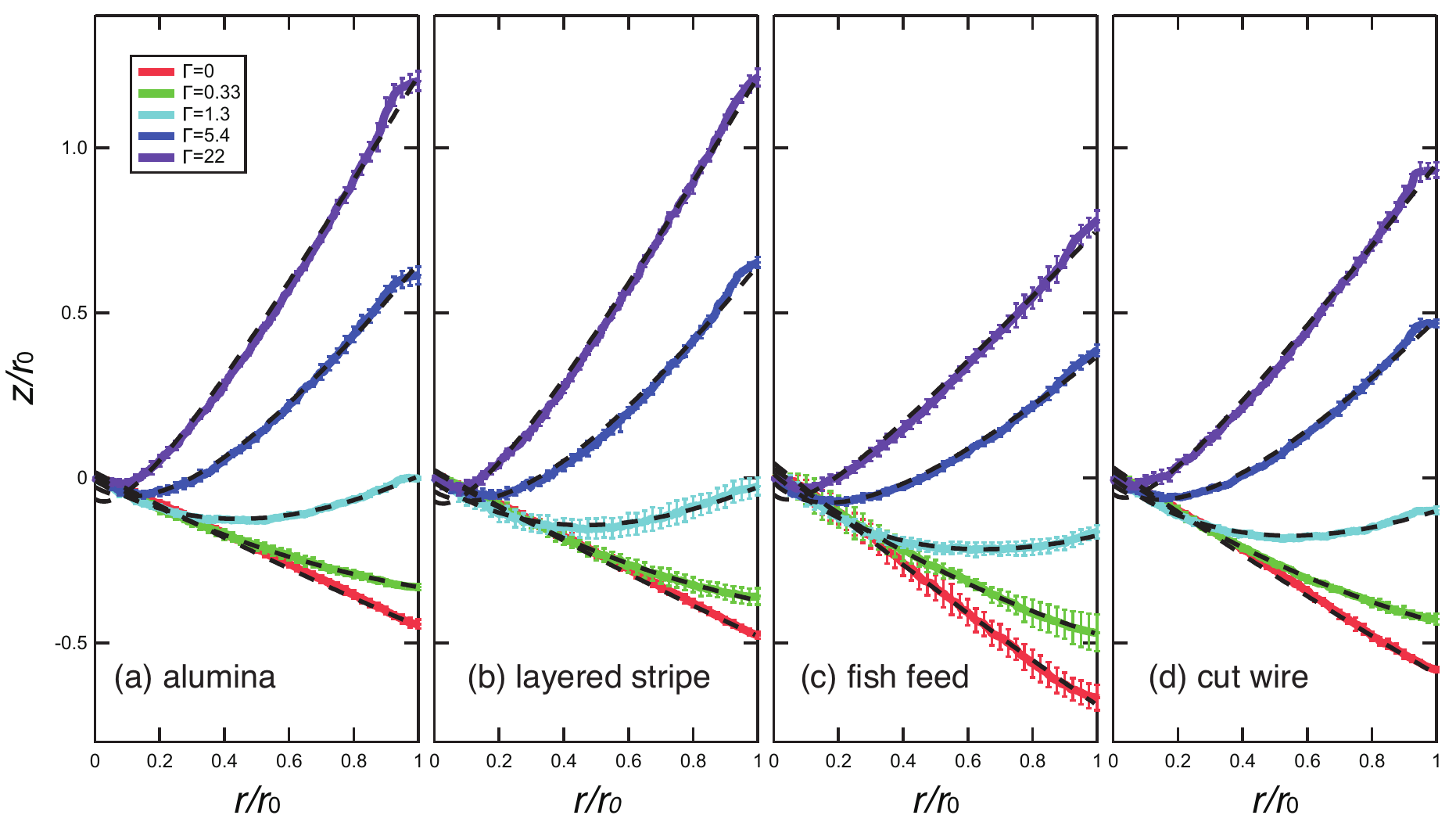}
\caption{
  Normalized surface profiles extracted by the image analysis of the raw data shown in Fig.~\ref{fig:raw_imgs} (in the increasing $\Gamma$ regime). The color indicates $\Gamma$ value as labeled in the legend. The dashed curves are the fitting by the force balance model of Eq.~(\ref{eq:model1}). Errors denote the standard deviation of the six (left and right of three experimental runs) profiles. 
}
  \label{fig:profiles_up}
\end{figure*}

To discuss the deformation mechanics, we extract the free-surface shape of the granular piles from the acquired images. The averaged (and normalized) surface profiles in the increasing $\Gamma$ regime are shown in Fig.~\ref{fig:profiles_up}. The normalized height $z/r_0$ is shown as a function of the normalized distance from the rotational axis $r/r_0$. To clearly show the deformation, the central peak position is fixed at the origin of the coordinate, i.e., $z(r=0)=0$. The colored solid curves and errors shown in Fig.~\ref{fig:profiles_up} are the average and standard deviation of six profiles, respectively. 

Although these profiles exhibit nontrivial curves, a simple model introduced in Ref.~\cite{Irie:2021b} can fit the experimental data very well. Dashed curves shown in Fig.~\ref{fig:profiles_up} represent the fitting by the model. In the next subsection, the force balance model used in the fitting is briefly explained.

\subsection{Force balance model in the increasing $\Gamma$ regime}
We consider the force balance among gravity, friction and centrifugal force at the surface of the granular pile. To achieve this force balance, the profile is quasi-statically deformed. At the equilibrium, the force balance (per unit mass) along the slope direction can be written as,
\begin{equation}
  g \sin \theta  + r \omega^2 \cos\theta = \mu \left( g\cos\theta - r \omega^2 \sin\theta \right),
  \label{eq:force_balance1}
\end{equation}
where $\mu$ and $\theta$ are the frictional coefficient and the local slope, respectively. It should be noted that $\mu$ does not represent the friction of grain contact. It rather characterizes bulk property of granular matter such as angle of repose. The clockwise direction of $\theta$ corresponds to the positive direction, and the range of $\theta$ is $-\pi/2 \leq \theta \leq \pi/2$. Then, by considering $dz/dr = -\tan\theta$, the force balance can be written as~\cite{Irie:2021b}, 
\begin{equation}
\frac{dz}{dr}= \frac{\Gamma r^* -\mu}{1+\mu \Gamma r^*},
\label{eq:model0}
\end{equation}
where $r^*=r/r_0$. An obvious relation, $dz/dr=-\tan\theta= -\mu$ is satisfied when $\Gamma=0$. Note that $\theta$ corresponds to the angle of local slope as schematically shown in Fig.~\ref{fig:setup}. This model is valid in both positive and negative $\theta$ cases. One can easily confirm that the form of Eq.~(\ref{eq:model0}) can be obtained also by substituting $\theta'=-\theta$ into Eq.~(\ref{eq:force_balance1}). Thus, the wide range of $\theta$ variation can be explained by Eq.~(\ref{eq:model0}) as far as $\Gamma$ is monotonically varied. By solving the differential equation (Eq.~(\ref{eq:model0})), solution for the surface profile can be obtained as~\cite{Irie:2021b},
\begin{equation}
  \frac{z}{r_0} =\frac{1}{\mu}\left[ r^* - \left( \frac{1}{\mu\Gamma} + \frac{\mu}{\Gamma} \right) \ln \left(\mu\Gamma r^* + 1 \right) \right] + C,  
  \label{eq:model1}
\end{equation}
where $C$ is an integration constant. 
Dashed curves shown in Fig.~\ref{fig:profiles_up} indicate the fitting by this form. In this model, $\mu$ and $C$ are free fitting parameters. We can clearly confirm that all the experimental data can be fitted very well by the model. This result is consistent with Ref.~\cite{Irie:2021b}. 

The value of $C$ relates only to the vertical level of the profiles. And, the fitted values are less than $O(10^{-2})$. On the other hand, $\mu$ is a crucial parameter characterizing granular friction. Therefore, we focus on the behavior of $\mu$ in the following. Specific values of $\mu$ and $C$ obtained by the fitting are listed in the supplementary materials.

In this model, we consider two-dimensional deformation of the granular pile. We neglect the effect of three dimensional deformation. While the deformation proceeds quasi-statically, inertial effect of the rotated grains could affect the deformation of the pile in the direction perpendicular to $r$-$z$ plane. Indeed, inertia-originated three-dimensional effect was confirmed in our previous study~\cite{Irie:2021b}. However, the effect of three-dimensional deformation is not very significant particularly for the estimate of $\mu$~\cite{Irie:2021b}. Thus, we follow the correction developed in~\cite{Irie:2021b} to consider the three-dimensional effect and simply use the two-dimensional model (Eq.~(\ref{eq:model1})).

\subsection{Friction in the increasing $\Gamma$ regime}

\begin{figure}
\centering
\includegraphics[clip,width=0.8\linewidth]{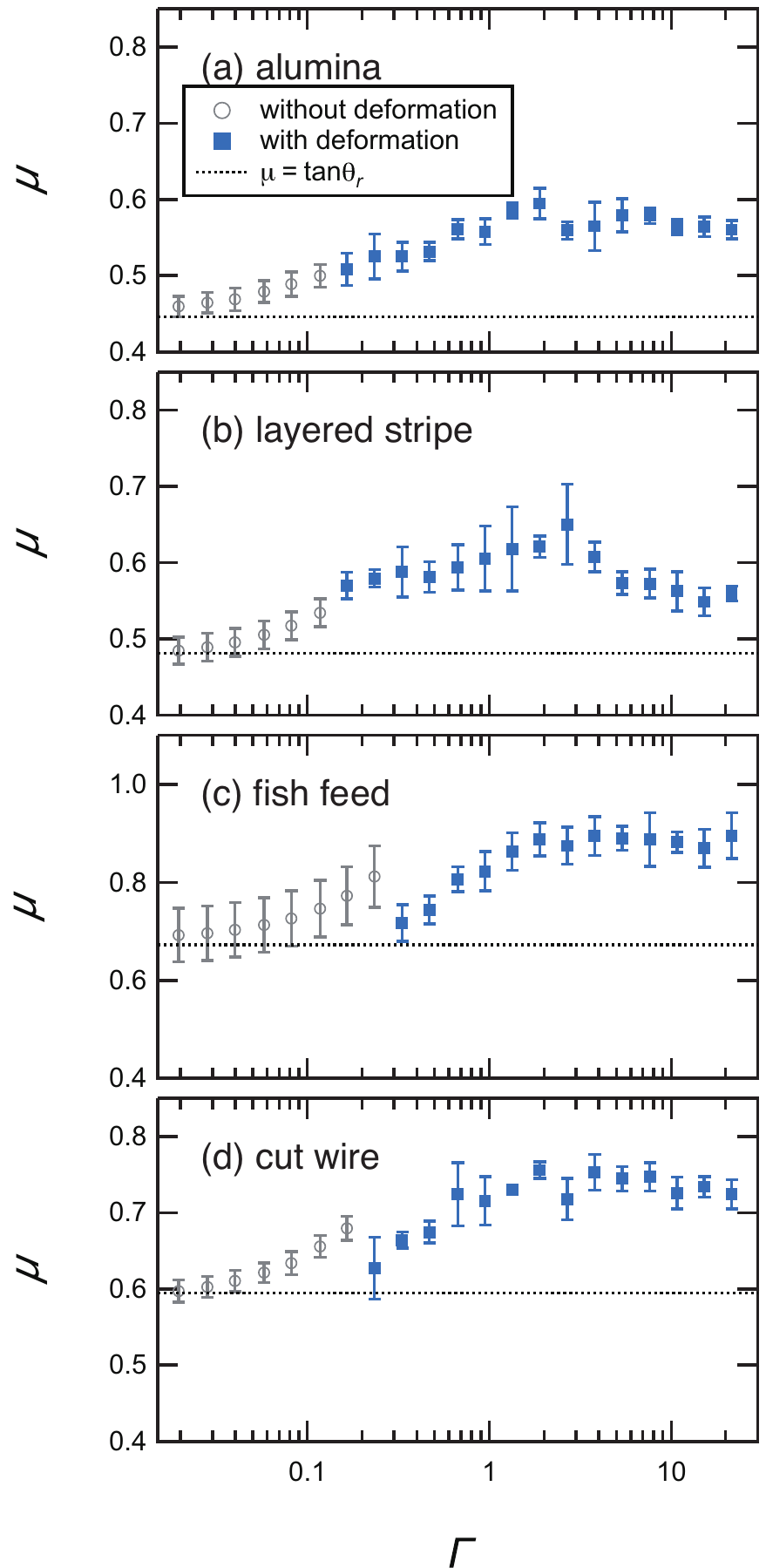}
\caption{
The relation between $\mu$ and $\Gamma$ in the increasing $\Gamma$ regime. The initial granular pile was not deformed in the gray circles region. Because the model assumes a finite deformation even in this regime, these data are not meaningful.  After the deformation was triggered, all $\mu(\Gamma)$ curves show the increasing trend (blue squares) at $\Gamma \lesssim 3$. Errors denote the standard deviation of the six (left and right of three experimental runs) profiles. The horizontal dotted lines indicate the level of the initial angle of repose, $\mu=\tan\theta_r$.
}
\label{fig:mu_G_up}
\end{figure}

In Fig.~\ref{fig:mu_G_up}, the fitted $\mu$ values in the increasing $\Gamma$ regime are presented. When $\Gamma \lesssim 0.2$, deformation of granular piles was not observed. This regime is displayed with gray circles in Fig.~\ref{fig:mu_G_up}. Although the increasing trend of $\mu(\Gamma)$ can be confirmed in this regime, this trend is an artifact. While the force balance model assumes a finite deformation of granular piles even in the small $\Gamma$ regime, the actual granular pile keeps its initial shape due to the small but finite cohesion effect originated from humidity effect, van der Waals force, electrostatic force, and so on. In other words, $\mu$ must become large to resist the sum of gravitational and centrifugal forces if we do not consider such cohesive forces. When $\Gamma=0$, $\mu$ is constant as far as the slope does not vary ($dz/dr=-\tan\theta=-\mu$ at $\Gamma=0$ in Eq.~(\ref{eq:model0})). However, when $\Gamma \neq 0$, $\mu$ depends on $\Gamma$ even when $\tan\theta$ is constant, as expressed in Eq.~(\ref{eq:model0}).
The increasing trend of $\mu$ in $\Gamma \lesssim 0.2$ comes from this effect. Using this property, we evaluated the effective cohesion strength in our previous paper~\cite{Irie:2021a}. Although we have considered the effective cohesion effect, this effect could result from other factors such as grain size effect that requires the lower limit of centrifugal force to trigger the grain motion. In this paper, we exclude this small $\Gamma$ part ($\Gamma \lesssim 0.2$) from the analysis to discuss the appropriate $\mu(\Gamma)$ behavior. As observed in the large $\Gamma$($\gtrsim 0.2$) regime, fitted $\mu$ values are greater than the initial values denoted by horizontal dashed lines in Fig.~\ref{fig:mu_G_up}. This tendency is consistent with our previous report~\cite{Irie:2021b}. Moreover, $\mu$ seems to be the maximum value around $\Gamma \simeq 3$. While grain density is widely varied in this study (from 0.61 to 4.57 ($\times 10^3$~kg~m$^{-3}$)), we confirm that the qualitative behaviors are similar in all grains. The principal question in this study is that the observed $\mu(\Gamma)$ trend is irreversible or not. To check the history dependence of $\mu(\Gamma)$ behavior, we analyzed the profiles in the decreasing $\Gamma$ regime as well.   

\subsection{Surface profiles in the decreasing $\Gamma$ regime}

\begin{figure*}
\centering
\includegraphics[clip,width=0.8\linewidth]{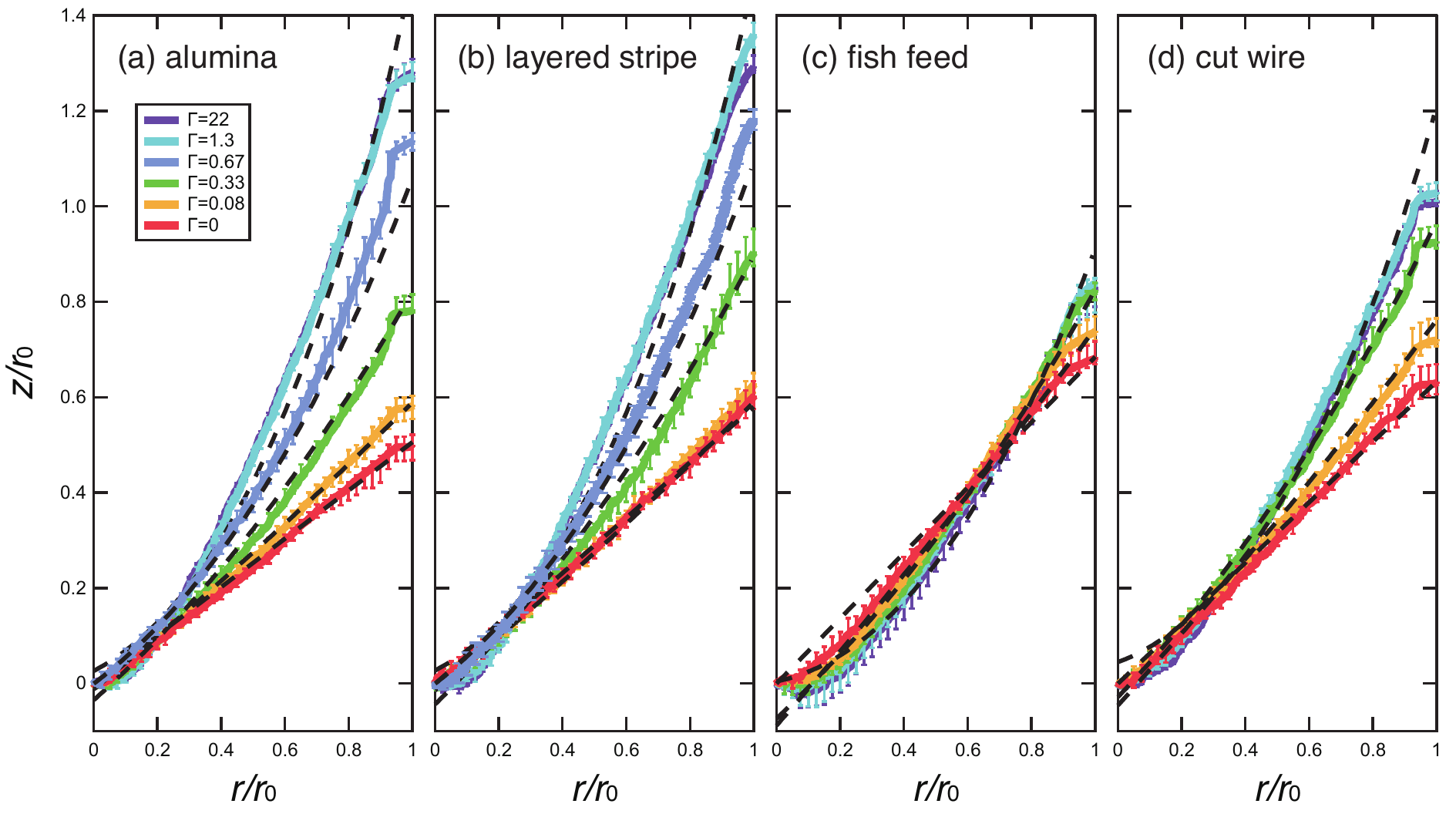}
\caption{
Normalized surface profiles in the decreasing $\Gamma$ regime are plotted. The color indicates $\Gamma$ value as labeled in the legend. The dashed curves are the fitting by the force balance model of Eq.~(\ref{eq:model3}) with free fitting parameters of $\mu$ and $C$. Errors denote the standard deviation of the six (left and right of three experimental runs) profiles. 
}
\label{fig:profiles_down}
\end{figure*}

In Fig.~\ref{fig:profiles_down}, observed surface profiles in the decreasing $\Gamma$ regime are presented. As mentioned before, granular piles cannot recover their initial shapes in spite of the same static state, $\Gamma=0$. More importantly, one can realize that the surface profiles produced by large $\Gamma$ are stable at least until $\Gamma$ is reduced down to $\Gamma \simeq 1$. When $\Gamma$ becomes less than the threshold, deformation occurs and finally the slightly relaxed piles are formed at the final state. As seen in Fig.~\ref{fig:profiles_down}, profiles made of fish-feed grains do not show significant deformation compared with alumina and layered-stripe cases. The deformation of cut-wire heaps show the intermediate behavior between them. These profiles should be related to the frictional property. 
To analyze the deformation of these profiles using the friction coefficient $\mu$, we have to slightly improve the force balance model.

\subsection{Force balance model in the decreasing $\Gamma$ regime}
In the decreasing $\Gamma$ regime, direction of surface flow becomes opposite. Thus, direction of friction also becomes opposite. We have to slightly modify the force-balance model. The angle $\theta$ of the profiles shown in Fig.~\ref{fig:profiles_down} is always negative by its definition (Fig.~\ref{fig:setup}). Therefore, we substitute the transformations, $\mu \to -\mu$ and $\theta \to - |\theta|$, into the force balance model of Eq.~(\ref{eq:force_balance1}). Then, from the geometrical setup, the collapse condition in the decreasing $\Gamma$ regime can be written as, 
\begin{equation}
  g \sin |\theta| - r \omega^2 \cos|\theta| \geq \mu \left( g\cos|\theta| + r \omega^2 \sin|\theta| \right).
  \label{eq:force_balance2}
\end{equation}
When this inequality is satisfied, deformation is induced.

This inequality explains the history dependence (strong stability) of the granular pile developed by large $\Gamma$. Specifically, Eq.~(\ref{eq:force_balance2}) can be rewritten as, 
\begin{equation}
  (1- \mu \Gamma r^*) \tan|\theta| \geq \mu + \Gamma r^*.
  \label{eq:force_balance3}
\end{equation}
In the decreasing $\Gamma$ regime, $\tan|\theta|$, $\mu$, and $\Gamma r^*$ are always positive. Thus, to hold the inequality, $1-\mu\Gamma r^*\geq 0$ must be satisfied. That is, deformation of granular piles at the sidewalls ($r^*=1$) can occur only when 
\begin{equation}
  \Gamma < \frac{1}{\mu}. 
  \label{eq:collapse_condition}
\end{equation}
This condition is consistent with the stability of the steep piles on the sidewalls in the decreasing $\Gamma$ regime. Once the deformation of the piles is triggered, the equilibrium shape is determined by the force balance again (the equality case of Eq.~(\ref{eq:force_balance2})). Then, the corresponding differential equation is obtained from the relation $dz/dr=\tan|\theta|$ as,
\begin{equation}
  \frac{dz}{dr}= \frac{\Gamma r^* +\mu}{1-\mu \Gamma r^*}.
  \label{eq:model2}
\end{equation}
When $\Gamma \geq 1/\mu$, the gradient $dz/dr$ diverges at $r^*=1/\mu\Gamma$. To avoid the divergence, the condition of Eq.~(\ref{eq:collapse_condition}) must be satisfied. 
The solution of Eq.~(\ref{eq:model2}) is computed as,
\begin{equation}
  \frac{z}{r_0} =\frac{1}{\mu}\left( r^* - \frac{1}{\mu\Gamma} \right) - \frac{1}{\Gamma}\left( 1+\frac{1}{\mu^2} \right) \ln \left( r^* -\frac{1}{\mu\Gamma}\right) + C.  
  \label{eq:model3}
\end{equation}
  
Dashed curves in Fig.~\ref{fig:profiles_down} show the fitting by this model. Although the quality of these fittings is not very good compared with the increasing $\Gamma$ regime, the curves capture the global trend of the deformation. Note that the complex effect of the three-dimensional deformation is neglected to derive Eq.~(\ref{eq:model3})  just like the derivation of Eq.~(\ref{eq:model1}). 

\subsection{Friction in the decreasing $\Gamma$ regime}

\begin{figure}
\centering
\includegraphics[clip,width=0.8\linewidth]{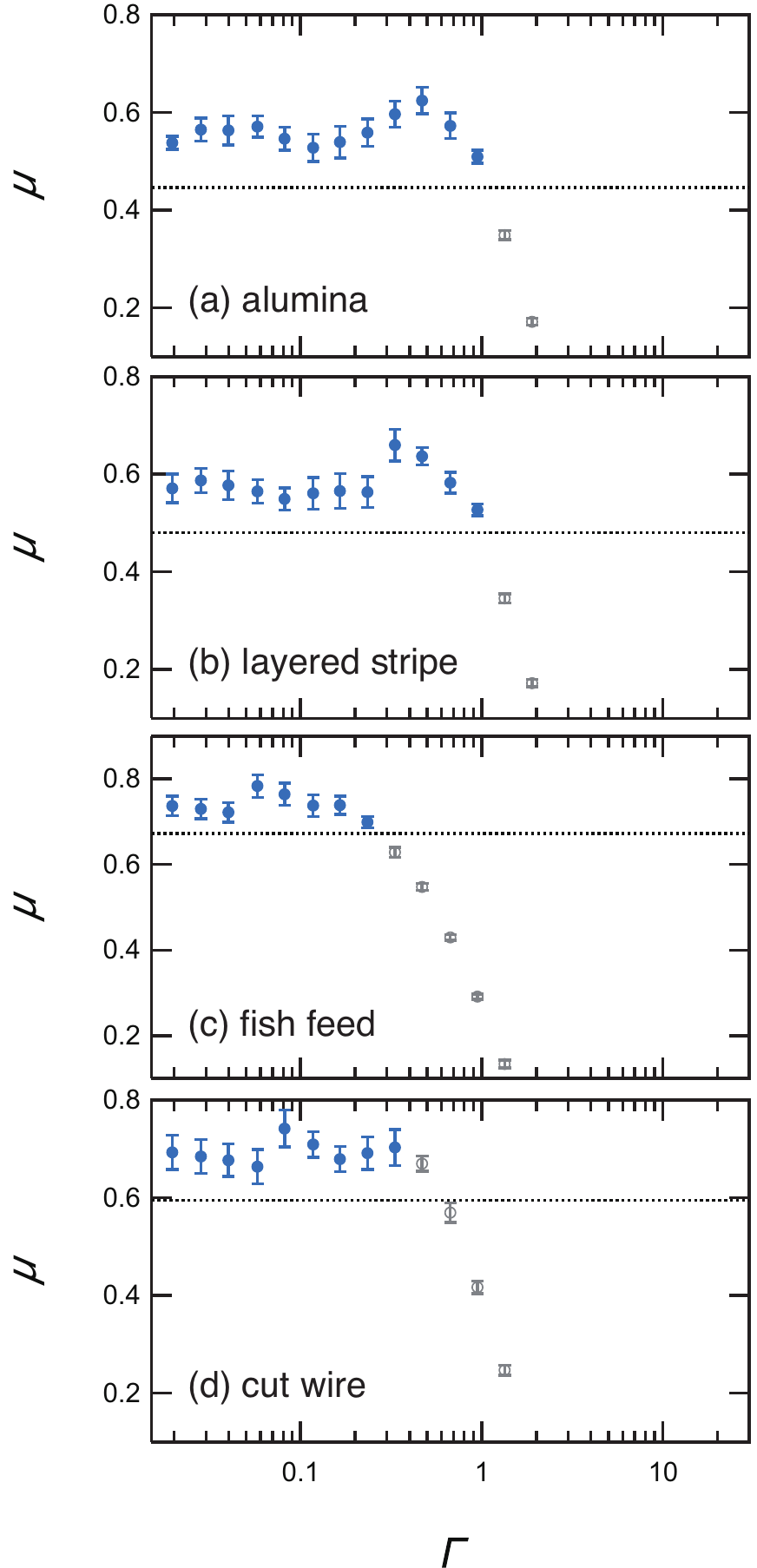}
\caption{
  The relation between $\mu$ and $\Gamma$ in the decreasing $\Gamma$ regime. Due to the model-applicability limit (Eq.~(\ref{eq:collapse_condition})), the fitting was performed when $\Gamma < 1/\mu$. Even in this model-applicable range, the obtained $\mu$ is very small in relatively large $\Gamma$ regime. The gray circles indicate that the granular-heap structure developed by the large $\Gamma$ did not deform. In small $\Gamma$ regime, $\mu$ becomes almost constant. Errors denote the standard deviation of the six (left and right of three experimental runs) profiles. The horizontal dotted lines indicate the level of the initial angle of repose, $\mu=\tan\theta_r$.
}
\label{fig:mu_G_down}
\end{figure}

In Fig.~\ref{fig:mu_G_down}, $\mu(\Gamma)$ relations in the decreasing $\Gamma$ regime are shown. The fitting result in the range of $\Gamma < 1/\mu$ is plotted. By the least-square fitting, we can somehow obtain the fitting curves in relatively large $\Gamma (\simeq 1)$ regime. However, the obtained $\mu$ value is unreasonably small in this regime. The actual granular piles at $\Gamma=1.3$ do not deform as shown in Fig.~\ref{fig:profiles_down}, whereas $1/\mu$ is greater than 1.3 for all cases. Indeed, too small $\mu$ values are estimated when the profiles developed by the large $\Gamma$ is kept unchanged~(Fig~\ref{fig:mu_G_down}). Namely, the criterion of Eq.~(\ref{eq:collapse_condition}) only provides the upper limit of $\Gamma$ above which the model cannot be applied. The model becomes truly applicable in smaller $\Gamma$ regime ($\Gamma \lesssim 0.1$). In such a small $\Gamma$ regime, the obtained $\mu$ values are almost independent of $\Gamma$ (Fig.~\ref{fig:mu_G_down}). Deformation of granular piles cannot be observed in relatively large $\Gamma$ regimes.

\section{Discussion}

\begin{figure}
\centering
\includegraphics[clip,width=0.8\linewidth]{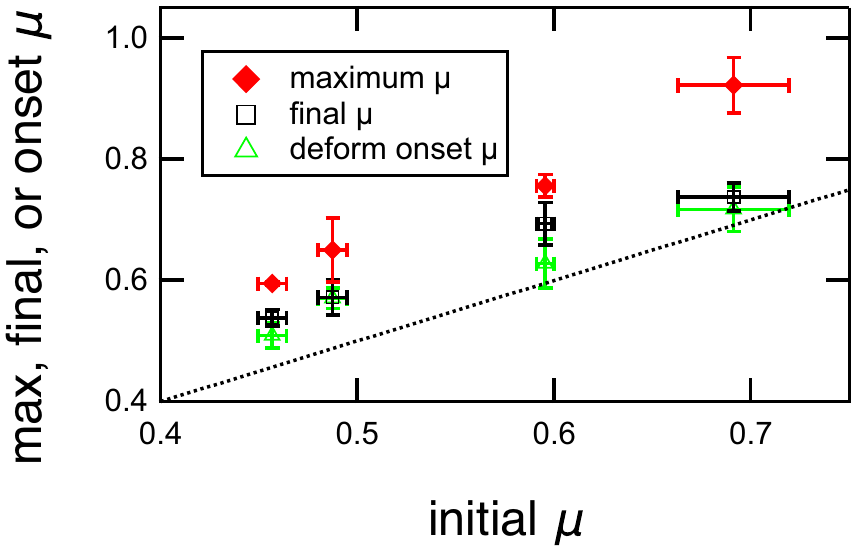}
\caption{
Comparison among the initial, maximum, final $\mu$ values, and $\mu$ at the deformation onset in the increasing $\Gamma$ regime. The maximum $\mu$ value is significantly greater than the initial $\mu$ values. The final $\mu$ values approach the middle values between the initial and the maximum $\mu$ values. The final values are close to the deformation onset values. The dotted line indicates the relation, (maximum, final, deformation-onset~$\mu$)~=~(initial~$\mu$). The data points from left to right correspond to the cases of alumina, layered stripe, cut wire, and fish feed, respectively.
}
\label{fig:ini_fin_mu}
\end{figure}

To clearly show the history-dependent behavior of $\mu$, relations among the initial $\mu$ ($\Gamma=0$), the maximum $\mu$, the final $\mu$ ($\Gamma=0$) values, and the $\mu$ value at which the deformation starts (in the increasing $\Gamma$ regime) are plotted in Fig.~\ref{fig:ini_fin_mu}. By applying the rotation, granular friction $\mu$ is increased. However, $\mu$ value reduces by decreasing $\Gamma$. As seen in Fig.~\ref{fig:ini_fin_mu}, the final $\mu$ values distribute between the initial and maximum $\mu$ values. This result clearly indicates the history-dependent $\mu$ behavior. Rather, the final $\mu$ values are close to the $\mu$ values at which the deformation starts in the increasing regime. This implies that the final $\mu$ values could be related to the starting angle (maximum stability angle) of granular system. In general, friction coefficient $\mu$ depends on various parameters such as state and rate variables even in quasistatic regime ~(e.g.~\cite{Fayed:1997} and \cite{Kawamura:2012}). To understand the complexly history-dependent behavior of $\mu$, further systematic investigation with other geometrical setup should also be performed for the comparison.

In this experiment, history dependence of $\mu$ is introduced by the combination of history-dependent deformation and change of the rotational acceleration sign. To precisely evaluate $\mu$ value in the increasing $\Gamma$ regime, the initial condition we employ is the best geometrical setup because that is a marginally stable state without rotation. Then, the deformation gradually proceeds by increasing $\Gamma$. This process inevitably results in two steep granular piles developed at both sides. Then, the rotational acceleration is switched from positive to negative. As a result, we observe a certain history-dependent deformation of the granular pile. It is difficult to independently vary the rotational acceleration and the shape of granular pile. The latter is determined by the rotational history. That is, the current experiment reveals the coupled effect of deformation history and rotational protocol. The word ``history dependence'' includes these two effects in this study. In this sense, the history dependence found in this study also means protocol dependence of granular behavior. Due to the complex history dependence, $\mu$ is not a unique function of $\Gamma$~(Figs.~\ref{fig:mu_G_up} and \ref{fig:mu_G_down}). However, the overlap of valid $\Gamma$ range in the increasing and decreasing $\Gamma$ regimes is very narrow. To ensure the history dependence of $\mu$, further investigation is necessary. 

According to the average behavior of the experimental results, the $\mu$ values approximately increase 33\% (at $\Gamma\simeq 3$) by applying the rotation. The $\mu$ value saturates or slightly decreases at $\Gamma \gtrsim 3$. If we assume the linear relation between $\mu$ and $\log_{10}\Gamma$ in the range of $\Gamma \lesssim 3$ (as roughly confirmed in Fig.~\ref{fig:mu_G_up}), we obtain a simple relation $\mu(\Gamma) = \mu_0(1+ c_1 \log_{10}\Gamma)$ (only in the increasing $\Gamma$ regime), where $\mu_0$ is the friction coefficient without centrifugal loading (initial value) and $c_1=0.69$ is a parameter obtained by the experimental result (33\% increase at $\Gamma=3$). Although this relation could be useful to roughly estimate the friction coefficient in the range of $0.1 \lesssim \Gamma \lesssim 3$, it should be noted that the relation is obtained based on the quasi-static deformation of the rotated granular pile in the increasing $\Gamma$ regime. In realistic applications, much more dynamic situations must also be considered. Thus, the frictional characterization in dynamically flowing regime must be studied in the future investigation. 

\begin{figure}
\centering
\includegraphics[clip,width=0.8\linewidth]{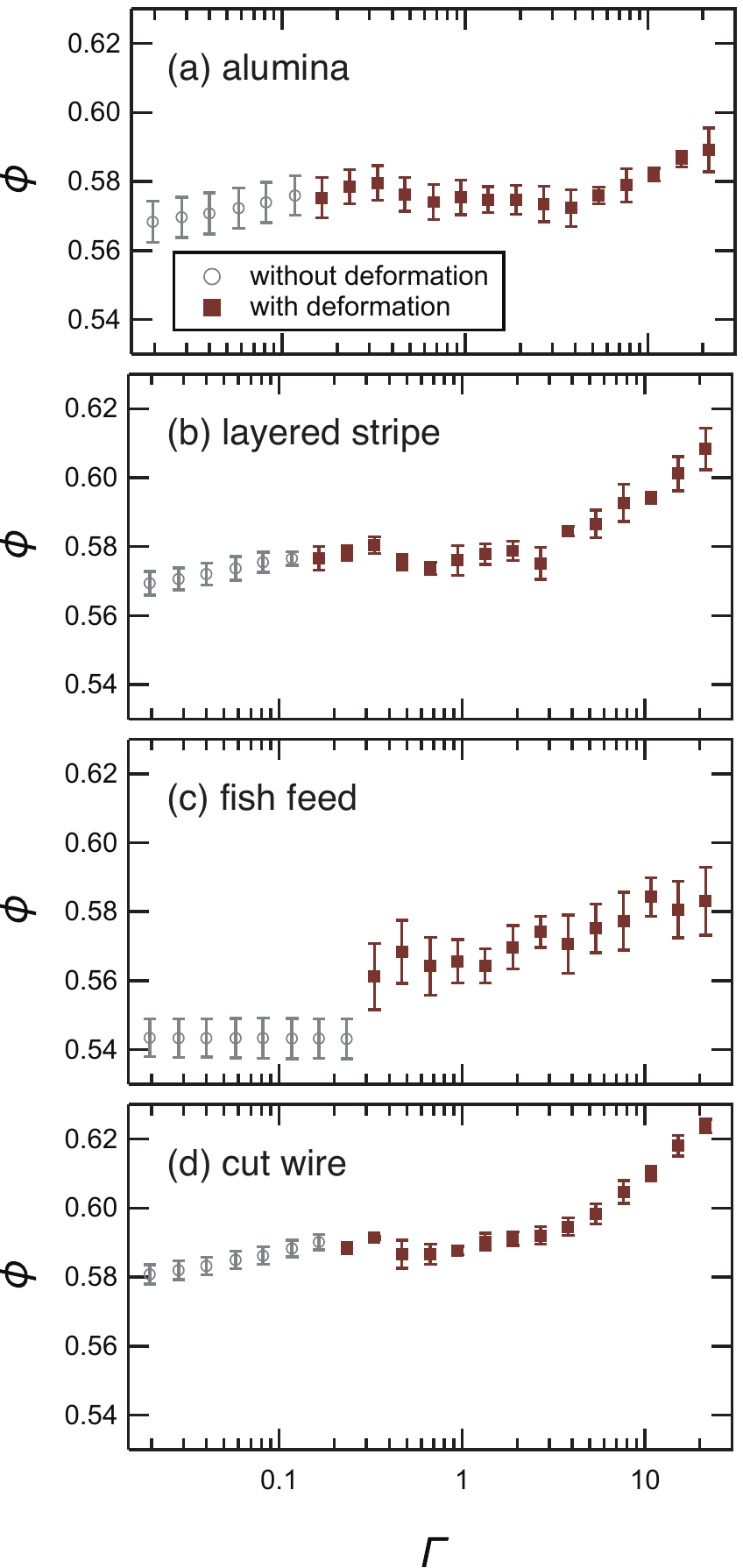}
\caption{
  The relation between $\phi$ and $\Gamma$ in the increasing $\Gamma$ regime. Gray symbols correspond to the no-deformation range. By increasing $\Gamma$, the granular pile is gradually compacted.  
}
\label{fig:phi_G_up}
\end{figure}

\begin{figure}
\centering
\includegraphics[clip,width=0.8\linewidth]{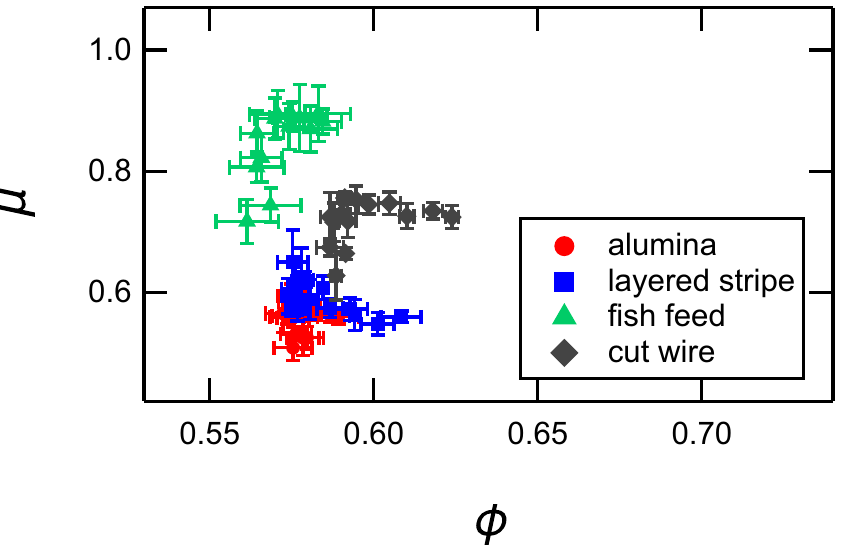}
\caption{
  The correlation between $\mu$ and $\phi$ in the increasing $\Gamma$ regime. The corner-like correlation can be observed. This correlation might suggest the altenate increase of $\mu$ and $\phi$ by the increase of $\Gamma$. 
}
\label{fig:mu_phi_corr}
\end{figure}

The average ratio between final $\mu$ and initial $\mu$ is $1.16$. This 16~\% increase is approximately the half of maximum increase (33~\%). For now, we are not sure about the universality of this relation. Because the $\mu$ value could depend on the network structure of the granular pile, microscopic characterization is necessary to further understand the origin of this tendency. As a macroscopic quantity which might relate to the $\mu$ behavior is the packing fraction~\cite{Fayed:1997}. Namely, the packing fraction could be a key factor to reveal the physical origin of $\mu$ variation.   

Therefore, we tried to measure the packing fraction of the granular pile. By applying additional body force (centrifugal force) to the granular pile, the grains network can be compressed. This compression might result in the increase in packing fraction and friction coefficient. We measured mass of grains in the cell and corresponding volume of the granular heap from the acquired images. From these data and the true density of constituent materials, we can estimate the packing fraction. 
Although the accuracy of the measured packing fraction is still limited, we carefully considered three-dimensional-deformation effect to evaluate the packing fraction. Details about the packing fraction measurement can be found in our previous paper~\cite{Irie:2021b}. In Fig.~\ref{fig:phi_G_up}, the measured packing fraction $\phi$ as a function of $\Gamma$ in its increasing regime is presented. As expected, granular compaction can be observed. Due to the technical limitation, measurement of $\phi$ in the decreasing $\Gamma$ regime was difficult. Specifically, surface three-dimensional-deformation effect in the decreasing $\Gamma$ regime cannot be surely measured. As discussed in Sec.~\ref{sec:profiles_increasing_G}, the effect of three-dimensional deformation does not significantly affect the estimate of $\mu$. However, $\phi$ is sensitive to the three-dimensional effect. Moreover, it is quite uncertain whether the correction method developed in the increasing $\Gamma$ regime can be applied also to the decreasing $\Gamma$ regime because the deformation of the rotated granular pile shows strong history dependence as we confirmed so far. Thus, we only focus on the increasing $\Gamma$ regime. Because $\phi$ must be a key quantity to characterize the variation in $\mu$, accurate measurement of $\phi$ even in the decreasing $\Gamma$ regime is an important future problem.

By comparing Figs.~\ref{fig:mu_G_up} and \ref{fig:phi_G_up}, one can realize all data show qualitatively similar behaviors in the increasing $\Gamma$ regime. When $\Gamma$ is small ($\Gamma \lesssim 0.1$--$0.2$), significant deformation cannot be detected. Thus, the increasing trend of $\mu$ in this regime is an artificial effect. However, the packing fraction $\phi$ in this regime shows a slightly increasing trend. This effect can be originated from the balance between centrifugal force and cohesion effect~\cite{Irie:2021a}. Then, in the intermediate regime ($0.1$--$0.2 \lesssim \Gamma \lesssim 1$--$3$), $\phi$ becomes almost constant and $\mu$ shows the increasing trend. In the large $\Gamma$ regime ($1$--$3 \lesssim \Gamma$), $\phi$ increases again and $\mu$ becomes roughly constant (or slightly decreasing). In this sense, rotation might result in the increase of $\phi$ or $\mu$. To clearly see the correlation, the relation between $\mu$ and $\phi$ is displayed in Fig.~\ref{fig:mu_phi_corr}. Indeed, the corner-like relation between $\mu$ and $\phi$ can be confirmed in Fig.~\ref{fig:mu_phi_corr}. However, this behavior suggests that $\mu$ cannot be simply controlled by $\phi$. To identify the principal mechanism increasing $\mu$, further investigation is necessary. Due to the technical limitation, accuracy of $\phi$ measurement is still limited and the obtained value represent the entire structure (not the local one)~\cite{Irie:2021b}. As mentioned in Sec.~\ref{sec:raw_images}, the deformation is localized in the surface thin layer. Thus, the local variation in $\phi$ should be related to the variation in $\mu$. However, the centrifugal force applied by the rotation affects all the bulk granular heap. Because the centrifugal force depends on $r$, $r$-dependent force gradient could also cause the variation in local $\phi$. Accurate measurement of local $\phi$ is necessary to further discuss this tendency. 

Finally, we briefly discuss the possible application of the methodology developed in this study. First, we can evaluate the centrifugal-force dependence of granular friction by using this experimental setup. Because the centrifugal loading is frequently used in various industrial situations (see \cite{Irie:2021a} and references therein), frictional characterization of the rotated granular matter will provide fundamental information necessary for efficiently hundling such centrifugal machines. At the same time, we can evaluate the effective cohesion strength of the granular matter by using the deformation threshold~\cite{Irie:2021a}. This property could be crucial for discussing slowly developing asteroidal surface deformation etc. To consider the specific application, we should perform additional experiments by using each corresponding target granular matter. In this paper, we build a basic methodology to establish the measuring method. The current experimental result we obtained in this study suggests that $\mu$ logarithmically depends on $\Gamma$. This means $\Gamma$ dependence of $\mu$ is weak. Therefore, for the industrial application, constant $\mu$ assumption might be sufficient in many cases.

\section{Conclusion}
In this study, we investigated the loading-history dependence of granular frictional coefficient $\mu$. Specifically, deformation of the surface profiles of the rotated granular pile was measured and analyzed to estimate $\mu$. For examining the history-dependent behavior, profiles in the increasing/decreasing $\Gamma$ regimes were systematically measured. As a consequence, we found that the profiles showed history-dependent deformation. Particularly, the granular piles developed by the rapid rotation (large $\Gamma$) were stably kept until the $\Gamma$ is reduced down to $\Gamma \simeq 1$. To understand this history dependence, we considered the force balance model both in increasing and decreasing $\Gamma$ regimes. Basically, we successfully reproduced the experimentally obtained profiles by the model. The measured $\mu$ shows an increasing trend in the regime of $0.1\lesssim \Gamma \lesssim 3$ and roughly saturates or slightly decreases at $\Gamma \gtrsim 3$. Besides, we found that the friction coefficient $\mu$ showed larger values than their initial values when the rotation is halted ($\Gamma=0$).  To estimate $\mu$ value, we obtained an empirical model, $\mu(\Gamma) = \mu_0(1+ + c_1 \log_{10}\Gamma)$ with $c_1=0.69$ in the range of $0.1 \lesssim \Gamma \lesssim 3$ only in the increasing $\Gamma$ regime. Besides, in the increasing $\Gamma$ regime, $\mu$ and $\phi$ seem to be alternately increased by adding the centrifugal loading. The evaluations of $\mu$ and effective cohesion strength using the developed experimental system would provide useful information for industry and planetary science. Because the current experiment only focussed on the quasi-static behavior, further study for dynamic characterization of $\mu(\Gamma)$ is a crucial next step.

\section*{Declaration of Competing Interest}
The authors declare that they have no known competing financial interests or personal relationships that could have appeared to influence the work reported in this paper.

\section*{Acknowledgment}
This work was supported by the JSPS KAKENHI, Grant No.~19H01951, 18H03679, and 17H06459.

\bibliographystyle{elsarticle-num} 
\bibliography{rot_relax}

\end{document}